\documentclass[twocolumn,superscriptaddress,pra,aps]{revtex4}
\usepackage{graphicx}
\usepackage{amsmath}
\usepackage{amssymb}
\usepackage{xcolor}

\begin{document}
\title{Effects of white noise on Bell theorem for qudits}
\author{Arijit Dutta}
\email{arijitdutta@kias.re.kr}
\affiliation{School of Computational Sciences, Korea Institute for Advanced Study, Seoul 02455, Korea}
\author{ Jaewan Kim}
\email{jaewan@kias.re.kr }
\affiliation{School of Computational Sciences, Korea Institute for Advanced Study, Seoul 02455, Korea}
\author{Jinhyoung Lee}
\email{hyoung@hanyang.ac.kr}
\affiliation{Department of Physics, Hanyang University, Seoul, 04763, Republic of Korea}

\newcommand{\Tr}{\mathrm{Tr}}
\newcommand{\bra}[1]{\left\langle #1\right|}
\newcommand{\ket}[1]{\left|#1\right\rangle}
\newcommand{\abs}[1]{\left|#1\right|}
\newcommand{\expt}[1]{\left\langle#1\right\rangle}
\newcommand{\braket}[2]{\left\langle{#1}|{#2}\right\rangle}
\newcommand{\commt}[2]{\left[{#1},{#2}\right]}

\newcommand{\ketbra}[2]{|#1\rangle \langle #2|}
\newcommand{\avg}[1]{\langle #1\rangle}

\begin{abstract}
We introduce two types of statistical quasi-separation between local observables to  construct two-party Bell-type inequalities for an arbitrary dimensional systems and arbitrary number of measurement settings per site. Note that, the main difference between statistical quasi-separations and  the usual statistical separations is that  the former  are not  symmetric under exchange of the  two local observables, whereas latter preserve the symmetry. We show that a variety of Bell inequalities can be derived by sequentially applying triangle inequalities which statistical quasi-separations satisfy. A sufficient condition is presented to show quantum violations of the Bell-type inequalities with infinitesimal values of critical visibility $v_c$. 
\end{abstract}

\maketitle

\section{introduction}

Bell theorem~\cite{Bell64} is one of the most profound quantum principle, particularly from the perspective of operationalism. It is intimate to the quantum entanglement, originating in the argument by Einstein, Podolsky, and Rosen in 1935~\cite{EPR35}. Bell theorem has unravelled the correlation nature, hidden in the randomness. It states that quantum theory is incompatible with local realistic theory as it violates the statistical inequalities of correlations, say Bell inequalities, that any local realistic theories obey. 

Noises generally prevent quantum systems from violating Bell inequalities~\cite{polozova2016higher}. Experimental devices (of preparation, operation, and/or measurement) suffer from noises as interacting with environments. As those devices need to be controlled by the outsider, it is difficult to completely eliminate the noises, even though some sophisticated method can reduce the amounts of noises~\cite{Wu2004}. It is thus an interesting question how much resilience the Bell violations show against noises. The stronger resilience against noises implies that Bell violation is still observable in a worse circumstance~\cite{Kaszlikowski00,CGLMP02}. One may focus on the type of white noise, as arises commonly on various occasions. For example, it arises by imperfect controls in experimental devices when the imperfections are isotropic~\cite{Nielsen00, loura2014}. Moreover, every colored noise is transformed to the white noise by the twirling operation~\cite{Werner89}.

Finding a Bell inequality is a hard task in general, in particular for a complex system with many subsystems, many settings of measurements, and many measurement outcomes. The methods include utilizing mathematical identities of discrete variables in the local realistic models~\cite{Bell64,CHSH69,Clauser74}, utilizing quantum identities in the reverse direction~\cite{lim2010genuinely}, finding facet inequalities in a local realistic polytope~\cite{Peres99}, and employing Kolmogorovian probabilities~\cite{santos1986,zohren2008}. In particular, the latter method has utilized triangle inequalities of correlations and developed to chained Bell-like inequalities with an arbitrary number of correlations for qubits~\cite{Pearle1970,pykacz1991} and qudits~\cite{Zukowski2014}. The chained inequalities have an interesting interpretation that the chained inequalities get close to a test of all versus nothing like Greenberger-Horne-Zeilinger theorem~\cite{GHZ89}.

The methodology of Kolmogorov probability theory is a powerful tool to derive Bell inequalities in a systematic manner. In Ref.~\cite{Dutta2018} the authors derived  a geometric extension of CH inequality for three sub-systems by introducing probability measure on symmetric difference  of probabilistic events.  Symmetry property of statistical separations is the key feature in derivation of such inequalities. It
was shown that due to the presence of symmetry  one can extend the formalism to derive Bell-type inequalities for arbitrary number of sites with arbitrary number of measurement settings per site.

Using a set of statistical quasi-separations between local observables, we derive  two-party Bell-type inequalities for  arbitrary dimensional systems and arbitrary number of measurement settings per site. Here ``quasi" symbolizes  the fact that the symmetry property is absent in quasi-separations. Due to the presence of asymmetry in quasi-separations of local observables, one needs to be careful about the ordering of local observables. The derivations of the inequalities are based on the sets of triangle inequalities. One of the most interesting features of the approach is that, one can show quantum violations of the inequalities against a purely random distribution, i.e., white noise with critical visibility $v_c\to 0$ (for details please see Eqs.~$\eqref{eq5c},$ and $\eqref{eq:curawn}$).
To this end, we show that a ratio of  quasi-separation of local observables for purely random distribution and the  algebraic upper bound of the quasi-separation must tend to zero. Note that, quantum violation of a Bell-type inequality for $v_c\to 0$ was also presented~\cite{Acin2002}, whereas any sufficient condition to see  such violations  was not provided. Ref.~\cite{Junge2010} employed the theory of operator algebra to show $v_c\to 0.$

Discussions on entanglement detections and Bell inequalities for qubits and higher dimensional systems based on noise models have been made~\cite{dutta2016, Gruca2012, khrennikov2008bell, laskowski2015highly, salavrakos2016bell, sen2003multiqubit, CGLMP02, vertesi2008more}.

The paper is arranged in the following way. In Sec.~\ref{sec:pidea}, we discuss the main idea of our work. A general derivation  of the inequalities and a sufficient  condition to see  quantum violations of the Bell-type inequalities for $v_c\to 0$ are discussed. In Sec.~\ref{whiten}, after introducing two types of quasi-separations, derivations and violations of Collins-Gisin-Linden-Massar-Popescu (CGLMP) inequality and a new two-setting Bell-type inequality for two quDits are presented. We also demonstrate how to show quantum violation of the new Bell-type inequality for $v_c\to 0.$ In Sec.~\ref{nset1}, we extend our formalism for many measurement settings per site. Precisely, we derive and check violations of both chained Bell-type inequalities with many measurement settings per site. We discuss the main results in Sec.~\ref{conclusion}.

\section{Principal idea}
\label{sec:pidea}
We present our principal idea to show the violations of Bell-type inequality against noises for infinitesimal values of critical visibility. To the end, we derive a set of Bell inequalities by employing a series of triangle inequalities that hold in any local realistic models. Such a Bell inequality is characterized by the quasi-distance, which is defined to satisfy triangle inequalities. We also derive a sufficient  condition that the quantum model violates a derived Bell inequality for $v_c\to 0.$


Consider an experiment where each of two observers Alice and Bob measures an observable, randomly chosen among $N$ observables. Their local observables are denoted by $A_{2n-1}$ and $B_{2n}$,  respectively, for $n=1, 2, \ldots, N$. We assume that every observable takes randomly one of $D$ integral outcomes from $0$ to $D -1$. 

We define `statistical' quasi-separation $S(A,B)$ between two local observables $A$ of Alice and $B$ of Bob in a form of
\begin{equation}
\label{eq:sqd}
\begin{aligned}
S(A, B) &= \sum_{a,b} d(a,b)\, p(a,b| A, B), \\
S(B, A) &= \sum_{a,b} d(b,a)\, p(a,b| A, B),
\end{aligned}
\end{equation}
where $p(a,b | A,B)$ is a probability distribution of outcomes $a$ and $b$ in the joint measurement of local observables $A$ and $B$ at Alice and Bob, respectively, and $d(a, b)$ is a quasi-distance of two outcomes $a$ and $b$. The quasi-separation means the average quasi-distance  how much separated the two local observables $A$ and $B$ are. The quasi-distance is not symmetric by permutation, $d(a,b) \ne d(b,a)$, and so is the quasi-separation $S$, i.e. $S(A,B) \ne S(B,A)$. This originates from the directionality that the quasi-distance of $a \rightarrow b$ is different from the one of $b \rightarrow a$. The directionality of distance is common in real life: For an example, the traveling time of distance depends on the way whether going up or down a hill. 

In local realistic models, the statistical quasi-separation $S(X,Y)$ satisfies the triangle inequality,
\begin{eqnarray}
\label{eq:sqdta}
S(X, Y) + S(Y,Z) \ge S(X,Z),
\end{eqnarray}
for every triple of local observables $(X, Y, Z)$. This is proved in Appendix \ref{sec:sqdlrm}. In quantum models, on the other hand, such a triangle inequality can not hold in general. These results remain valid for every quasi-distance $d(x,y)$, if `points' $x$ and $y$ are mapped to the outcomes of local observables $X$ and $Y$ with no particular form of quasi-distance. 

The approach of triangle principle derives a Bell-type inequality by adding two or more triangle inequalities. A triangle inequality includes quasi-separations of local observables at a given observer, say $S(A_i,A_j)$, which needs to be removed for a Bell-type inequality. Such quasi-separations can be cancelled out by adding another triangle inequality. We adopt this method to derive a Bell-type inequality. This simple method provides options to derive a set of Bell-type inequalities. One example is a chained Bell-type inequality of many local observables. 
Here, we present some examples of deriving Bell-type inequalities by the method.

For example, consider a triangle inequality for the triple of local observables $(A_1, B_2, A_3)$, given by
\begin{eqnarray}
  \label{eq2}
S(A_1,B_2)+S(B_2,A_3) - S(A_1,A_3) \ge 0.
\end{eqnarray}
The quasi-separation $S(A_1,A_3)$ with the negative sign is cancelled out by adding another triangle inequality with $S(A_1,A_3)$ positively signed, i.e.,
\begin{eqnarray}
 \label{eq3}
S(A_1,A_3)+S(A_3,B_4) - S(A_1,B_4) \ge 0.
\end{eqnarray}
Adding these two triangle inequalities result in a quadrangle inequality,
\begin{eqnarray}
 \label{eq:quadineq}
S(A_1,B_2) + S(B_2,A_3) + S(A_3,B_4) - S(A_1,B_4) \ge 0.
\end{eqnarray}
This is called a closed form of Bell-type inequality with four local observables, excluding any quasi-separations $S(X_n, X_m)$ of observables at a given observer $X$. In fact, this closed form is led to  CGLMP inequality with a particular quasi-distance $d(x,y) = x-y \mod D =: [x-y]_D$, where $[x-y]_D$ is a positive residue modulo $D$~\cite{acin2005,CGLMP02}. Adding more triangle inequalities can derive a chained Bell inequality of $2N$ local observables,
\begin{multline}
\label{chain0}
\sum_{n=1}^{N-1} \left( S(A_{2n-1},B_{2n}) + S(B_{2n},A_{2n+1}) \right) \\
+ S(A_{2N-1},B_{2N}) - S(A_1, B_{2N}) \ge 0.
\end{multline}
Another option is adding the closed forms with four local observables, as in Eq.~\eqref{eq:quadineq}, resulting in a Bell inequality of $2N = 4M$ local observables,
\begin{eqnarray}
\label{eq:sof4lo}
\sum_{m=1}^M {\cal B}_m \ge 0,
\end{eqnarray}
where ${\cal B}_m= S(A_{4m-3},B_{4m-2}) + S(B_{4m-2},A_{4m-1}) + S(A_{4m-1},B_{4m}) - S(A_{4m-3},B_{4m})$. There are more options to derive Bell-type inequalities. One of them can be  to extend our work to a multipartite system by chaining local observables of quasi-separations between pairs of subsystems. We do not pursue, however, in this direction.

Let us focus on a sufficient condition to show violation of Bell-type inequality against white noise for $v_c\to 0.$ Consider a Bell function $I_N$ with $2N$ local observables, defined by the left hand side as in Eq.~\eqref{eq:quadineq}, Eq.~\eqref{chain0}, or Eq.~\eqref{eq:sof4lo}, so that the Bell-type inequality is given by
\begin{eqnarray}
\label{eq55}
  I_N \ge 0,
\end{eqnarray}
where $I_C = 0$ is the classical lower bound in local realistic models. 
Note that we consider an arbitrary form of Bell-type inequality in the approach of triangle principle.

 Let $N_T$ be the number of triangle inequalities to add up.  Note that, ${\cal B}_m\geq 0$ and each of them can be obtained by adding two triangle inequalities similar to Eqs.~$\eqref{eq2}$ and $\eqref{eq3}.$ When we add two triangle inequalities, quasi-separation $S(A_{4m-3},A_{4m-1})$ is cancelled out. Thus, if we consider the Bell-type inequality~$\eqref{eq:sof4lo}$ is obtained by summing up $M$ closed forms, then $2M$ number of quasi-separations are cancelled out. Also, we assume $I_N$ involves $N_S$ quasi-separations with $N_+$ ones positively-signed and $N_-$ ones negatively-signed, where $N_S = 3N_T - 2M$, $N_+ = 2N_T-M$, $N_- = N_T - M$, and $N_S = N_+ + N_-$.

Let the Bell-type inequality $\eqref{eq55}$ be violated by quantum expectations with a pure quantum state $\hat{\rho} = |\psi\rangle \langle \psi |$. Consider a quantum state, which is an output state decohered in a channel of white noise,
\begin{eqnarray}
  \label{eq5a}
   \hat{\rho}_v = v \hat{\rho} + (1-v) \hat{\rho}_{r}, 
 \end{eqnarray}
 where $v,$ the fraction of the pure quantum state $\hat{\rho},$ is defined as visibility and $\hat{\rho}_{r}=\openone \otimes \openone/D^2$ is the purely random state of white noise. Note that $0\le v\le 1$.
Assume that $I_N$ is a linear function of $ \hat{\rho}_v$. Then, one obtains
\begin{eqnarray}
\label{eq5b}
I_N\left( \hat{\rho}_v \right) = v I_N\left( \hat{\rho} \right) + (1-v) I_N\left( \hat{\rho}_{r} \right),
\end{eqnarray}
where $ I_N\left( \hat{\rho} \right)$ and $I_N\left( \hat{\rho}_{r} \right)$ are quantum expectations of $I_N$ by $\hat{\rho}$ and $\hat{\rho}_{r}$, respectively. The critical visibility $v_c$ is defined such that, for $v > v_c$, $\hat{\rho}_{v}$ violates the Bell-type inequality $\eqref{eq55}$, i.e.,  $I_N\left( \hat{\rho}_{v} \right) <0.$
By the condition $I_N\left( \hat{\rho}_{v_c} \right) = I_C=0$, the critical visibility $v_c$ can be quantified as
\begin{eqnarray}
\label{eq5c}
v_c= {{I_N\left( \hat{\rho}_{r} \right)} \over {I_N\left( \hat{\rho}_{r} \right)-I_N\left( \hat{\rho} \right)}} .
\end{eqnarray}
Note that, if the critical visibility $v_c$ is small, then the violation is more resilient against the noise. In particular, if $v_c \to 0$,  the Bell-type inequality is violated even by an infinitesimal fraction of the pure state $\hat{\rho}$ with $v>v_c$. The condition $v_c \to 0$ is rephrased by $I_N\left( \hat{\rho}_{r} \right)/\left|I_N\left( \hat{\rho} \right) \right| \to 0$:
\begin{eqnarray}
\label{eq:curawn}
v_c \to 0 \Longleftrightarrow \frac{I_N\left( \hat{\rho}_{r} \right)}{\left|I_N\left( \hat{\rho} \right) \right|} \to 0.
\end{eqnarray}

The Bell function $I_N$ consists of quasi-separations with positive and negative signs, as in Eq.~\eqref{chain0} or \eqref{eq:sof4lo}. $I_N$ decreases by decreasing (increasing) the quasi-separations with the positive (negative) signs. We take this idea for $I_N(\hat{\rho})$ of the pure state $\hat{\rho}$. Let $S_{\mathrm{max}}$ be the algebraic upper bound of $S(X,Y)$ so that $0\le S(X, Y)\le S_{\mathrm{max}}, \forall (X,Y)$. The quasi-separation $S(X,Y)$ is scaled with respect to $S_\text{max}$: $S(X,Y) = (S_\text{max}/2) \left( 1 + s(X,Y) \right)$, where $-1 \le s(X,Y) \le 1$. Assume that, for the pure state $\hat{\rho}$, $I_N(\hat{\rho})$ is negatively valued by quantum expectations of quasi-separations with the same values (in an appropriate choice of local measurements). Precisely, the quasi-separations with $\pm$ signs have the same values of $S_\pm = (S_\text{max}/2) \left( 1 \mp s_\pm \right)$, respectively, where $0 \le s_\pm \le 1$. 
This does not necessarily lead to the optimal Bell violation~\footnote{Nonetheless, there are some known cases that Bell violation is even optimal. The quantum expectation of Bell function equivalent to the Clauser-Horne-Shimoney-Holt of 2 qubits is given by $I_2 = S_{12}+S_{23}+S_{34}-S_{14} = 3 S_+ - S_-=1-\sqrt{2} < 0$, where $S_{ij} = S(A_i, B_j)$ and  $S_\pm=(1 / 2) \left(1 \mp (1 /\sqrt{2})\right)$. This is the optimal violation.}. 
Then, $I_N\left( \hat{\rho} \right)$ is given by
\begin{eqnarray}
\label{eq6}
I_N\left( \hat{\rho} \right) &=& N_+ S_+ - N_- S_-\nonumber\\
&=& - S_\text{max} \, \xi,
\end{eqnarray}
where $2 \xi = N_- \left( 1 + s_-\right)  - N_+ \left( 1 - s_+ \right)$. The  violation is indicated by $\xi > 0$, or equivalently, $( 1 - s_+ )/( 1 + s_-) < N_- / N_+$. For the random distribution $ \hat{\rho}_{r}$, on the other hand, the pair joint probability $p(a,b|A,B) = 1/D^2$ so that all quasi-separations are the same, $S(A,B) = S(B,A) = (1/D^2) \sum_{x,y} d(x,y) =: S_r$ for every pair of local observables $A$ and $B$. We then obtain
\begin{eqnarray}
\label{eq66}
I_N\left( \hat{\rho}_r \right) &=& (N_+ - N_-) S_{r}.
\end{eqnarray}
Combining Eqs.~\eqref{eq6} and \eqref{eq66}, we obtain
\begin{eqnarray}
\frac{I_N \left( \hat{\rho}_{r} \right)}{\left| I_N\left( \hat{\rho} \right) \right|} = \frac{1}{\zeta} \frac{ S_{r}}{S_\mathrm{max}},
\end{eqnarray}
where $\zeta =\left|\xi \right|/(N_+-N_-)$ is always finite. Thus, the sufficient condition for $v_c\to 0$ in  Eq.~\eqref{eq:curawn} is attained when 
\begin{eqnarray}
\label{eq:ncfur}
\mathrm{SSR} := {S_{r} \over S_\mathrm{max}} \to 0,
\end{eqnarray}
where SSR stands for the scaled quasi-separation for the purely random distribution. We provide examples that, if condition $\eqref{eq:ncfur}$ holds, one obtains quantum violation with critical visibility $v_c \to 0$. Here the optimal violation is not crucial. 

\section{Quantum violations with four settings}
\label{whiten}

By employing the Bell-type inequality, as in Eq.~\eqref{chain0}, we investigate the quantum violation and its resilience against white noise. For the purpose, we employ a pure state $\hat{\rho} = |\psi \rangle \langle \psi|$, and a set of local measurements represented by Fourier bases~\cite{zeilinger1993einstein, CGLMP02, Son06a}, which were suggested to generalize a beam splitter into more ports and called a ``tritter'' for three input/output ports~\cite{zeilinger1993einstein}. In particular we test whether a quasi-separation satisfies the sufficient condition in Eq.~\eqref{eq:ncfur}, and compare the test to the tendency of violation. Two types of quasi-separations are examined in this work.

\subsection{Types of quasi-separations}
As seen in Eq.~\eqref{eq:sqd}, our quasi-separation $S(X,Y)$ is defined by the statistical average of a quasi-distance $d(x,y)$. In this work we employ a quasi-distance in a form of
\begin{eqnarray}
\label{eq:qdcglmp1}
d(x,y) = \left[ f(x) - f(y) \right]_D,
\end{eqnarray}
where $[x]_D = x \mod D$ is congruent to a nonnegative residue modulo $D$ for integral $x$, so that it belongs to the set $\{0, 1, ..., D-1\}$. Here, $f(x)$ is an integer-valued function. Appendix~\ref{sec:qdfbi} proves that $d(x,y)$ in Eq.~\eqref{eq:qdcglmp1} is a quasi-distance with an arbitrary function $f(x)$. Thus, function $f(x)$ defines the quasi-distance $d(x,y)$ and eventually the quasi-separation $S(X,Y)$ in our method.

We consider two types of $f(x)$, one of a linear function defined by
\begin{eqnarray}
\label{eq:lfinnr}
f_\text{I}(x) = x
\end{eqnarray}
and the other of a nonlinear function,
\begin{eqnarray}
\label{eq:nlfinnr}
f_\text{II}(x) = \delta_{x,0},
\end{eqnarray}
where $\delta_{x,y}$ is the Kronecker delta. The first type $f_\text{I}(x)$ has intensively been utilized to derive CGLMP inequality of two qudits~\cite{CGLMP02,acin2005} and Svetlichny-like inequality of more quDits~\cite{bancal2011detecting}. On the other hand, the second type $f_\text{II}(x)$ has not been employed yet to our knowledge. 

For both types, the $S_r$ and $\text{SSR}$ in Eq.~\eqref{eq:ncfur} are obtained in Appendix~\ref{sec:qswn}: For $f(x) = f_\text{I}(x)$ in Eq.~\eqref{eq:lfinnr},
\begin{eqnarray}
\label{eq:srssrt1}
S_r = \frac{D-1}{2} \quad \text{and} \quad \text{SSR} = \frac{S_r}{S_\text{max}} = \frac{1}{2}.
\end{eqnarray}
This SSR is just a constant, depending on no parameters. In other words, the sufficient condition in Eq.~\eqref{eq:ncfur} can not be fulfilled so that the quantum violation cannot be obtained for $v_c\to 0.$ For $f(x) = f_\text{II}(x)$ in Eq.~\eqref{eq:nlfinnr}, on the other hand,
\begin{eqnarray}
\label{eq:srssrt2}
S_r = \frac{D-1}{D} \quad \text{and} \quad \text{SSR} = \frac{S_r}{S_\text{max}} = \frac{1}{D}.
\end{eqnarray}
(see Appendix~\ref{sec:qswn}).
It is seen that $\text{SSR} \to 0$ as $D \to \infty$. These results suggest that the nonlinear quasi-distance $d(x,y)$ with $f(x) = f_\text{II}(x)$ is a candidate for the quantum violation against white noise for $v_c\to 0.$

\subsection{CGLMP inequality: Type I}
\label{rank}

In this subsection, we discuss derivation and quantum violation of CGLMP inequality for two quDits. When the linear function $f(x) = f_\text{I}(x)=x$ is employed,  the quasi-distance $d(x,y) = \left[x-y\right]_D$ in Eq.~\eqref{eq:qdcglmp1} and then the quadrangle inequality in Eq.~\eqref{eq:quadineq} becomes CGLMP inequality (see Refs.~\cite{acin2005}),
\begin{equation}
\label{eq:cglmp}
\begin{aligned}
I_2 = \, & S_\text{I}(A_1,B_2) + S_\text{I}(B_2,A_3) \\
& + S_\text{I}(A_3,B_4) - S_\text{I}(A_1,B_4),
\end{aligned}
\end{equation}
where $S_\text{I}$ are quasi-separations defined with type-I linear function $f(x)=f_\text{I}(x)$.
The CGLMP is said type-I, as defined with the type-I function, $f_\text{I}(x)$. To obtain the usual form of the CGLMP inequality~\cite{CGLMP02}, quasi-separations $S_\text{I}(A_i,B_j) $ and  $S_\text{I}(B_j, A_k)$ are replaced with scaled-down quasi-separations, i.e., $S_\text{I}(A_i,B_j) $ and  $S_\text{I}(B_j, A_k)$ are multiplied by $2/(D-1).$ Using the definition of quasi-separation (see Appendix~\ref{sec:sqdlrm} for explanation) one can write
\begin{eqnarray}
\label{eq:00}
\begin{aligned}
&\frac{2}{D-1}S_\text{I}(A_i,B_j)=\frac{2}{D-1}\sum_{a_i, b_j=0}^{D-1}[a_i-b_j]_Dp(a_i, b_j)&\\
&=1+\sum_{k=0}^{\lfloor D/2\rfloor-1}(1-\frac{2k}{D-1})(p(a_i=b_j-k-1)&\\&-p(a_i=b_j+k)),&
\end{aligned}
\end{eqnarray}
and
\begin{eqnarray}
\label{eq:11}
\begin{aligned}
&\frac{2}{D-1}S_\text{I}(B_j, A_l)=\frac{2}{D-1}\sum_{b_j, a_l=0}^{D-1}[b_j-a_l]_Dp(a_l, b_j)\\
&=1+\sum_{k=0}^{\lfloor D/2\rfloor-1}(1-\frac{2k}{D-1})(p(b_j=a_l-k-1)&\\&-p(b_j=a_l+k)).&
\end{aligned}
\end{eqnarray}
Inserting $\eqref{eq:00}$ and $\eqref{eq:11}$ in $\eqref{eq:cglmp}$, we can retrieve the usual form of  the CGLMP inequality~\cite{CGLMP02}.
In this case, $\text{SSR} = 1/2$ with $S_\text{max} = D-1$ and $S_r = (D-1)/2$, as shown in Eq.~\eqref{eq:srssrt1}.  {\em As the sufficient condition~\eqref{eq:ncfur} does not hold, one cannot see quantum violation of the CGLMP inequality against the white noise for $v_c\to 0.$}

We present the quantum violation of the CGLMP inequality in terms of the critical visibility $v_c$ in Eq.~\eqref{eq5c}, as in Ref.~\cite{CGLMP02}, before we compare to the other type of Bell inequality.
Consider a noisy quantum state $\hat{\rho }_v$ of an output in a white-noisy channel, as in Eq.~$\eqref{eq5a}$. The input state $\hat{\rho} = |\psi\rangle \langle \psi|$ is assumed maximally entangled with Schmidt rank $R=D$, i.e. 
\begin{eqnarray}
\label{eq:maxes}
|\psi\rangle =\frac{1}{\sqrt{D}}\sum_{i=0}^{D-1}|j j \rangle,
\end{eqnarray}
where $\{|j\rangle\}$ is a standard basis.
The entangled state $|\psi\rangle$ violates CGLMP inequality with local measurements of $A_n$ and $B_m$, represented by Fourier bases,
\begin{eqnarray}
\label{eq:lmba}
\ket{a}_n &=&
\frac{1}{\sqrt{D}}\sum_{j=0}^{D-1} \omega_D^{j (a +\alpha_n)}|j \rangle, \\
\label{eq:lmbb}
\ket{b}_m &=&
\frac{1}{\sqrt{D}}\sum_{j=0}^{D-1} \omega_D^{-j (b +\beta_m)}|j \rangle,
\end{eqnarray}
where $\omega_D = \exp \left( 2i\pi / D \right)$ is a primitive $D$-th root of unity, and constants $\alpha_1=0, \beta_2=-1/4, \alpha_3=-1/2$, and $\beta_4=-3/4$. Then, the joint probability $p(a, b|A_n, B_m)$ that Alice and Bob respectively obtain outcomes $a$ and $b$ in their local measurements of $A_n$ and $B_m$, is given by
$p_{nm}(a, b) := p(a, b|A_n, B_m) = \Tr \left( \hat{\rho} \, \hat{A}_n(a) \otimes \hat{B}_m(b) \right)$, where the projectors $\hat{A}_n(a) = |a\rangle_{n n} \langle a|$ and $\hat{B}_m(b) = |b\rangle_{m m}\langle b|$. Then,
\begin{align}
\label{probcglmp}
p_{nm}(a, b) &= 
\frac{g_1^{nm}(0)}{g_D^{n m}(a-b)},
\end{align}
where  $g_X^{n m}(x)= X^3 \sin^2\left[(\pi/X) (x+\alpha_n-\beta_m)\right]$. 
The quasi-separations $S_\text{I}(A_n, B_m)$ of the positive sign are shown to be equal to each other in Appendix~\ref{sec:qbfti} for $N=2$, and then
\begin{equation}
\label{eq:qstimc}
\begin{aligned}
S_+	&= S_\text{I}(A_1, B_2) = S_\text{I}(B_2,A_3) = S_\text{I}(A_3,B_4) \\
	&= \bar{C}_D, \\
S_- 	&= S_\text{I}(A_1, B_4) =  D-1 - \bar{C}_D,
\end{aligned}
\end{equation}
where $\bar{C}_D = \sum_{c=0}^{D-1} c \, p_D(c)$ is the average of correlation and $p_D(c)$ is the marginal probability of correlation, $p_D(c) = 1/2 D^2 \sin^2\left[ \pi (c+1/4)/D \right]$ (see Appendix~\ref{sec:qpampc}).
Then the Bell function $I_{2q}:=I_2(\hat{\rho})$ of CGLMP is given by Eq.~\eqref{eq6},
\begin{equation}
\begin{aligned}
I_{2q} &= - S_\text{max} \, \xi = - (D-1) \left( 1 - \frac{4 \bar{C}_D}{D-1} \right).
\end{aligned}
\end{equation}
For white noise $\hat{\rho}_r = \openone \otimes \openone/D^2$, on the other hand, 
\begin{align}
\label{eq:probcglmpwn}
p_{nm}(a, b) &= \frac{1}{D^2},
\end{align}
the quasi-separations $S_\text{I}(A_n,B_m) = S_r = (D-1)/2$, and the Bell function $I_{2r}:=I_2(\hat{\rho}_r)$ is given by Eq.~\eqref{eq66},
\begin{align}
\label{eq:probcglmprbf}
I_{2r} = D-1.
\end{align}
The critical visibility $v_c$ is obtained by Eq.~\eqref{eq5c},
\begin{align}
v_c = \frac{1}{1 + \xi},
\end{align}
where $\xi = 1 - 4 \bar{C}_D/(D-1)$.

Table~\ref{tab:cglmp} presents the numerical values of $I_{2q}$, $I_{2r}$, and $v_c$ for a couple of $D$. Note that $v_c$ decreases as increasing $D$. As $D \to \infty$, however,
\begin{eqnarray}
\label{eq:acvcglmp}
v_c \to \pi^2/\left( 16 \, \text{Catalan} \right) \simeq  0.67344,
\end{eqnarray}
where Catalan's constant $\text{Catalan} \simeq 0.91597$, as in Ref.~\cite{CGLMP02}. Thus, the critical visibility $v_c$ is bounded below by the finite value $\simeq  0.67344$, as reflected by $\text{SSR} = 1/2$.

\begin{table}[t]
\caption{Bell functions $I_{2q}$ for the quantum state in Eq.~\eqref{eq:maxes} of Schmidt rank $R=D$, $I_{2r}$ for white noise, and critical visibilities $v_\text{c}$. 
}
\centering
\begin{ruledtabular}
\begin{tabular}{l c  c   c   c  }
Bell inequality & $D$	& $I_{2q}$				& $I_{2r}$			& $v_c$ \\
\hline
type I		& $2$	& $-0.4142$			& 1.0000			& 0.7071	\\	
of CGLMP	& $3$     	& $-0.8729$			& 2.0000			& 0.6962	\\
		& $4$  	& $-1.3444$			& 3.0000			& 0.6906	\\
		& $5$  	& $-1.8211$			& 4.0000			& 0.6872	\\
		& $6$  	& $-2.3005$			& 5.0000			& 0.6849	\\
\hline
type II	& $2$	& $-0.4142$			& 1.0000			& 0.7071	\\	
		& $3$     	& $-0.3769$			& 1.3333			& 0.7796	\\
		& $4$  	& $-0.3620$			& 1.5000			& 0.8056	\\
		& $5$  	& $-0.3548$			& 1.6000			& 0.8185	\\
		& $6$  	& $-0.3508$			& 1.6667			& 0.8261	\\
\end{tabular}
\end{ruledtabular}
\label{tab:cglmp}
\end{table}

\subsection{Type-II Bell-type inequality}
\label{typeII}
We consider the type-II Bell inequality of two quDits, which is defined in Eq.~\eqref{eq:quadineq} with the nonlinear type-II function, $f(x) = f_\text{II}(x)=\delta_{x,0}$. The type-II Bell-type inequality is the same as the CGLMP of type I, except the type of function $f(x)$ [see Eqs.~\eqref{eq:lfinnr} and \eqref{eq:nlfinnr}]. Note that $\text{SSR} = 1/D$, as shown in Eq.~\eqref{eq:srssrt2} (or Appendix~\ref{sec:qswn}). {\em This implies that $\text{SSR} \to 0$ in the limit of $D \to \infty$, so that the type-II Bell inequality satisfies the sufficient condition~\eqref{eq:ncfur} for quantum violation against white noise for $v_c\to 0.$} However, the entangled state shows the weaker resilience of quantum violation, i.e., $v_c$ increases,  with $D$ than the type I of CGLMP, if the Schmidt rank $R$ is approximately equal to the dimension $D$ of white noise. For $R \ll D$, on the other hand, an entangled state the critical visibility $v_c \to 0$ as $D \to \infty$.

%
The quasi-separations of type II are given, from Eqs.~\eqref{eq:sqd}, \eqref{eq:qdcglmp1}, and~\eqref{eq:nlfinnr}, in forms of
\begin{eqnarray*}
\label{non1111}
S_\text{II}(A_n, B_m)&=& (D-1)\sum_{a=1}^{D-1}p_{nm}(a, 0) + \sum_{b=1}^{D-1}p_{nm}(0,b), \\
\label{non2222}
S_\text{II}(B_m, A_n)&=& \sum_{a=1}^{D-1}p_{nm}(a, 0) + (D-1)\sum_{b=1}^{D-1}p_{nm}(0,b),
\end{eqnarray*}
where $p_{nm}(a,b) = p(a,b|A_n,B_m)$.
The Bell function $I_2$ is given by the left hand side in Eq.~\eqref{eq:quadineq}, 
\begin{equation}
\begin{aligned}
I_2 = & \, S_\text{II}(A_1,B_2) + S_\text{II}(B_2,A_3) \\
&+ S_\text{II}(A_3,B_4) - S_\text{II}(A_1,B_4).
\end{aligned}
\end{equation}

For the state of Schmidt rank $R=D$ in Eq.~\eqref{eq:maxes} and the measurement bases in Eqs.~\eqref{eq:lmba} and \eqref{eq:lmbb} [or equivalently, the joint probabilities in Eq.~\eqref{probcglmp}], the quasi-separations $S_\text{II}(A_n, B_m)$ with the positive sign are shown to be equal to each other in Appendix~\ref{sec:qbftii} for $N=2$ and $R=D$. Then,
\begin{equation}
\begin{aligned}
S_+ 	& = S_\text{II}(A_1, B_2) = S_\text{II}(B_2,A_3) = S_\text{II}(A_3,B_4) \\
	&= 1- p_D(0), \\
S_- 	&= S_\text{II}(A_1, B_4) = 1- p_D(-1),
\end{aligned}
\end{equation}
where $p_D(c) = 1/2 D^2 \sin^2\left[ \pi (c+1/4)/D \right]$ is the marginal probability of correlation.
Then the Bell function $I_{2q}:=I_2(\hat{\rho})$ of type II is given by Eq.~\eqref{eq6},
\begin{equation}
\label{eq:i2qtiiqd}
\begin{aligned}
I_{2q}	&= - \left( D - 1 \right) \, \xi, \\
\xi		&= \frac{1}{D-1} \left[ \left(1-p_D(-1)\right) - 3 \left(1-p_D(0)\right) \right].
\end{aligned}
\end{equation}
For white noise $\hat{\rho}_r = \openone \otimes \openone/D^2$, on the other hand, the quasi-separations $S_\text{II}(A_n,B_m) = S_r = (D-1)/D$, and the Bell function $I_{2r}:=I_2(\hat{\rho}_r)$ is given by Eq.~\eqref{eq66},
\begin{align}
\label{eq:i2qtiird}
I_{2r} = \frac{2(D-1)}{D}.
\end{align}
The critical visibility $v_c$ is given by Eq.~\eqref{eq5c},
\begin{eqnarray}
\label{eq:qcvtiifr}
v_c = \frac{1}{1 + D \, \xi/2}.
\end{eqnarray}

Table~\ref{tab:cglmp} presents the numerical values of $I_{2q}$, $I_{2r}$, and $v_c$ for a couple of $D$. Note that $v_c$ increases as increasing $D$, implying that the resilience of quantum violation gets worse with $D$. In other words, the type I of CGLMP inequality shows stronger resilience against white noise than the type II for the state of Schmidt rank $R=D$. A related work was reported for CGLMP inequality~\cite{polozova2016higher}: The resilience of violation against white noise decreases with the dimension $D$ of Hilbert space, when  the fraction of the pure state is given by the $D$-th power of $v$. 

\subsection{Noise effects on type II}
\label{sec:urtii}

When Schmidt rank $R$ of a given pure state is much smaller than the dimension $D$ of white noise, the resilience of violation becomes stronger. We investigate the resilience of quantum violation for Schmidt rank $R < D$ of a pure state. 

We employ the state and the local measurements, the same as those in the previous subsections, except the rank $R \ne D$. The input state $\hat{\rho} = |\psi\rangle \langle \psi|$ is assumed entangled with Schmidt rank $R < D$, i.e. 
\begin{eqnarray}
\label{eq:maxesrr}
|\psi\rangle =\frac{1}{\sqrt{R}}\sum_{j=0}^{R-1}|j j \rangle,
\end{eqnarray}
where $\{|j\rangle\}$ is a standard basis.
The local measurement bases $\{|a\rangle_n\}$ of $A_n$ are given by
\begin{equation}
\label{eq:lmbarr}
\ket{a}_n =
\left\{
\begin{aligned}
& \frac{1}{\sqrt{R}}\sum_{j=0}^{R-1} \omega_R^{j (a +\alpha_n)}|j \rangle && \text{if $a < R$} \\
& |a\rangle && \text{otherwise},
\end{aligned}
\right.
\end{equation}
where $\omega_R = \exp \left( 2i\pi / R \right)$ is a primitive $R$-th root of unity, and constants $\alpha_1=0$ and $\alpha_3=-1/2$. 
Similary, the measurement bases $\{|b\rangle_m\}$ of $B_m$ are given by 
\begin{eqnarray}
\label{eq:lmbbrr}
\ket{b}_m =
\left\{
\begin{aligned}
& \frac{1}{\sqrt{R}}\sum_{j=0}^{R-1} \omega_R^{-j (b +\beta_m)}|j \rangle && \text{if $b < R$} \\
& |b\rangle && \text{otherwise},
\end{aligned}
\right.
\end{eqnarray}
where constants $\beta_2=-1/4$ and $\beta_4=-3/4$. Then, the joint probability $p_{nm}(a, b) := p(a, b|A_n, B_m)$ is given by
\begin{eqnarray}
\label{probcglmprr}
p_{nm}(a, b) = 
\left\{ 
\begin{aligned}
& \frac{g_1^{n m}(0)}{g_R^{n m}(a-b)} && \text{if $a,b<R$} \\
&0 && \text{otherwise}
\end{aligned}
\right.
\end{eqnarray}
where  $g_R^{n m}(x)= R^3 \sin^2\left[(\pi/R) (x+\alpha_n-\beta_m)\right]$. Then, we obtain the Bell function $I_{2q} := I_2(\hat{\rho})$ by Eq.~\eqref{eq6} (see Appendix~\ref{sec:qbftii}),
\begin{equation}
\label{eq:qbftii}
\begin{aligned}
I_{2q} &= - D \, \xi'_R \\
\xi'_R &= \frac{1}{R} \left[ \left(1-p_R(-1)\right) - 3 \left(1-p_{R}(0)\right) \right], 
\end{aligned}
\end{equation}
where $p_R(c) = 1/2 R^2 \sin^2\left[ \pi (c+1/4)/R \right]$. Here, we scaled $I_{2q}$ by $D$ instead of $S_\text{max}$, so that $\xi'_R$ does not depend on $D$. As $I_{2r} = 2(D-1)/D$ as in Eq.~\eqref{eq:i2qtiird}, the critical visibility $v_c$ is given by
\begin{eqnarray}
\label{eq:qcvtii}
v_c = \frac{1}{1 + \frac{D^2}{2(D-1)} \, \xi'_R}.
\end{eqnarray}
Note that $\xi'_R$ in Eq.~\eqref{eq:qbftii} does not depend on $D$. {\em As $D \to \infty$, thus, $v_c \to 0.$ This remains valid for finite Schmidt ranks $R$ of entangled pure states.} This is one of our main results. Note that, in our derivation we have considered  states $\eqref{eq:maxesrr}$ with Schmidt rank $R<D.$ Thus, implicitly, we have assumed that for such states $(D-R)$ Schmidt coefficients are zero. Interestingly, violations of Bell inequalities certify device independent scenario in communication protocols. More explicitly, in such cases communicating  parties do not trust the source which emits entangled states and also measurement setups. Thus, the certification of device independence entirely depends on the measurement statistics. However, in this work while considering critical visibility as a measure of resilience of the violations of Bell-type inequalities against white noise, one has to know the dimension of the emitted state. Otherwise, when $D\to \infty,$ effectively, the noise model does not have any effect on the measurements which are restricted to finite dimensions. Violations of the type-II Bell-type inequality~\ref{typeII} for finite values of $R$ and $D\to\infty$ leads to $v_c\to 0.$ Note that, if the violations of  the Bell-type inequalities is required to certify security of a device independent quantum key distribution (QKD) protocol, then the information about the dimension of the state makes the protocol semi device independent. The protocol is unsecure as an adversary may have access to the source and can manipulate the dimension of the state. Thus, in such a case the critical visibility may not be a good quantity to certify security of the QKD protocol.

\begin{figure}[t]
\begin{center}
\includegraphics[width=0.45\textwidth]{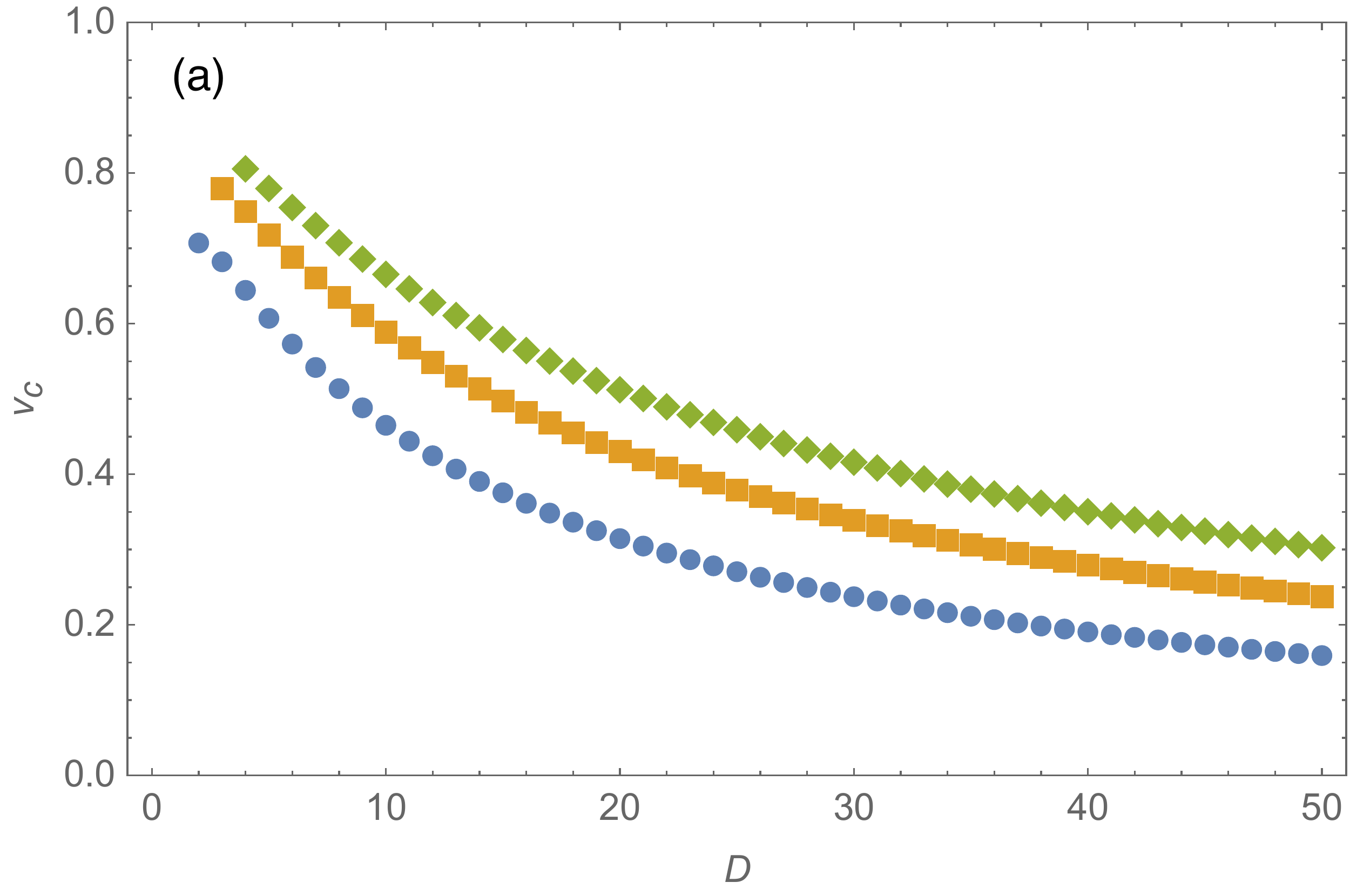} \\
(a) \\ 
\includegraphics[width=0.45\textwidth]{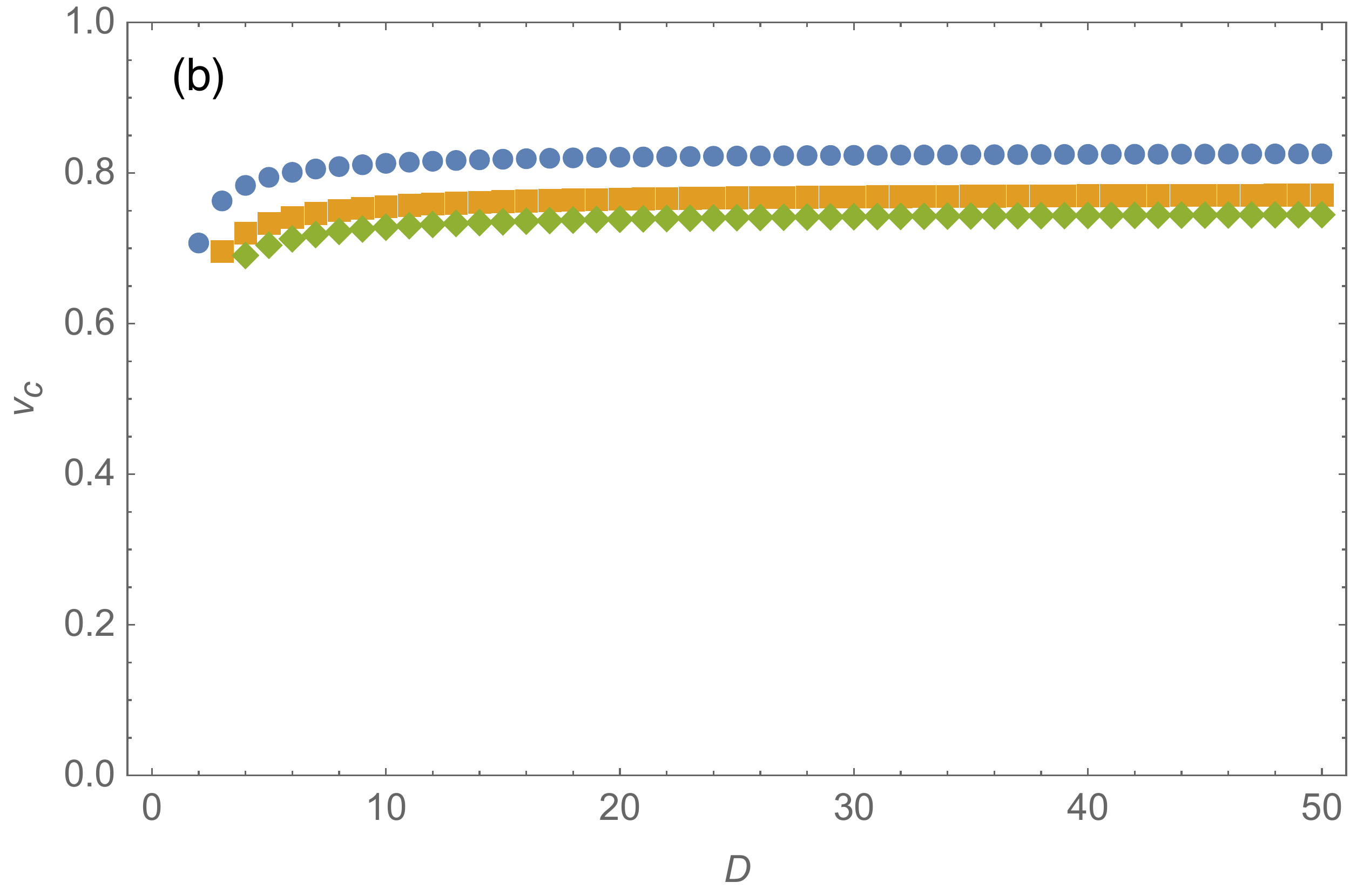} \\
(b)
\caption{(color online) Critical visibility $v_c$ as a function of the dimension $D$ of white noise for Schmidt ranks $R=2,3,$ and $4$ (circle, square, and diamond, respectively). (a) For the type II inequality, the $v_c$ decreases to zero with $D$ for each $R$. For given $D$, $v_c$ is the lowest with $R=2$ and increases with $R$. These tendencies are opposite to the type I of CGLMP, as shown at (b).}
\label{fig:1}
\end{center}
\end{figure}

Fig.~\ref{fig:1} presents the critical visibility $v_c$ as a function of the dimension $D$ for typical Schmidt ranks $R=2,3,4$ (circle, square, diamond). For the type II, Fig.~\ref{fig:1}(a) shows that the $v_c$ decreases to zero with $D$ for each $R$. For given $D$, $v_c$ is the lowest with $R=2$ and increases with $R$. As a reference, the results for the type I of CGLMP are presented in Fig.~\ref{fig:1}(b); they show the tendencies opposite to the type II. 

The similar results to the type II were discussed in Ref.~\cite{Acin2002} for Clauser-Horne-Shimoney-Holt (CHSH) inequality \cite{CHSH69} with the entangled state of Schmidt rank $R=2$. However, our work is not restricted to the CHSH inequality nor the lowest Schmidt rank(s), and by the triangle principle it opens an avenue to generalize Bell-type inequalities with arbitrary quasi-distances and to quantum violations for $v_c\to 0.$ Also, in Ref.~\cite{Junge2010}, using the formalism of operator space theory authors presented that for $v_c\to 0,$ one can obtain violations of Bell-type inequalities. Although, the authors considered all local realistic probabilities as noise models but they did not specify any set of quantum states and measurements which would lead to such condition. Instead, the derivation was based on the existence of quantum probabilities which lead to violations of Bell-type inequalities for $v_c\to 0.$

\section{Chained inequalities with many settings}
\label{nset1}
One can extend the analysis to $N$ local observables per site with the chained inequality $I_N \ge 0$, where $I_N$ is given by the left hand side in Eq.~\eqref{chain0}. Consider CGLMP-like chained inequality of type I. The algebraic lower bound of $I_N$ is given by 
\begin{align}
\label{eq:cialb}
I_N \ge - S_\text{max} = - (D-1).
\end{align}
As shown in Appendix~\ref{sec:qbfti} for general $N$, the quantum Bell function $I_{Nq} := I_N(\hat{\rho})$ is given by Eq.~\eqref{eq6},
\begin{eqnarray}
\begin{aligned}
I_{Nq} 		&= - (D-1) \, \xi, \\
\xi 	&= 1 - \frac{2N}{D-1} \bar{C}_D.
\end{aligned}
\end{eqnarray}
where $\bar{C}_D = \sum_{c=0}^{D-1} c \, p_D(c)$ is the average of correlation and $p_D(c)$ is the marginal probability of correlation, $p_D(c) = \sin^2\left( \pi/2N \right)/D^2 \sin^2\left[ \pi (c+1/2N)/D \right]$. Note $p_D(c)$ is not the same as the one below Eq.~\eqref{eq:qstimc} but depends on the number $N$ of local observables (see Appendix~\ref{sec:qpampc}). In the limit of $N \gg 1$, $\xi$ is approximated upto the first order of $1/N$,
\begin{eqnarray}
\begin{aligned}
\xi 
&\approx 1 - \frac{1}{N} \left[ \frac{\pi^2}{2(D-1)} \sum_{x=1}^{D-1} \frac{x}{D^2 \sin^2\left( \pi x / D\right)} \right].
\end{aligned}
\end{eqnarray}
Thus, $I_{Nq}$ approaches the algebraic lower bound of $- S_\text{max}$ as $N \to \infty$. In other words, as increasing the number of local observables, the Bell test is close to the kind of ``all or nothing'' like Greenberger-Horne-Zeilinger theorem \cite{GHZ89}. This is compared to the similar conclusion in Ref.~\cite{Zukowski2014}, where each basis of a local measurement is constructed by sub-dimensional unitary operations on a standard basis. 

For general $N$, the critical visibility
\begin{align}
v_c = \frac{1}{1 + \xi / (N-1)}, 
\end{align}
where we used $I_{Nr} = (N-1)(D-1)$. As $N \to \infty$, $v_c \to 1$. This implies that the quantum violation of the CGLMP-like chained inequality of type I becomes fragile by the white noise as the number of local observables are increased. There is no advantage in increasing the number of local observables, even though the quantum violation approaches the algebraic lower bound. These results also hold in the type-II chained inequality.

Consider the type-II chained inequality. The algebraic lower bound of $I_N$ is the same as Eq.~\eqref{eq:cialb}.
As shown in Appendix~\ref{sec:qbftii} for general $(N, D, R)$, the quantum Bell function $I_{Nq} := I_N(\hat{\rho})$ is given by Eq.~\eqref{eq6},
\begin{equation}
\label{eq:qbftiin}
\begin{aligned}
I_{Nq} 	&= - D \, \xi'_R, \\
\xi'_R		&= \frac{1}{R} \left[ \left(1-p_R(-1)\right) - (2N-1) \left(1-p_{R}(0)\right) \right],
\end{aligned}
\end{equation}
where $p_R(c) = \sin^2\left( \pi/2N \right) / R^2 \sin^2\left[ (\pi/R) (c+1/2N) \right]$ (see Appendix~\ref{sec:qpampc}). Here, we scaled $I_{2q}$ by $D$ instead of $S_\text{max}$ similarly to Eq.~\eqref{eq:qbftii} and $\xi'_R$ does not depend on $D$.  In the limit of $N \gg 1$, $\xi'_R$ is approximated upto the first order of $1/N$,
\begin{equation}
\begin{aligned}
\xi'_R \approx \, & \frac{1}{R} + \frac{\pi^3}{2} \sum_{x=1}^{R-1} \frac{\cos\left(\pi x/R\right)}{R^4 \sin^3\left(\pi x/R\right)} \\
& - \frac{1}{N} \left( \frac{\pi^2}{2} \sum_{x=1}^{R-1}\frac{1}{R^3 \sin^2\left(\pi x/X\right)} \right).
\end{aligned}
\end{equation}
The $\xi'_R$ decreases with $R$ with the maximum $\xi_2'$ and $\xi'_{2} \to 1/2$ as $N \to \infty$, so that the algebraic lower bound is not attainable. For general $(N, D, R)$, the critical visibility $v_c$ is given as 
\begin{eqnarray}
\label{eq:qcvtiin}
v_c = \frac{1}{1 + \frac{D^2}{2(D-1)} \frac{1}{N-1} \, \xi'_R},
\end{eqnarray}
where we used $I_{Nr} = 2(N-1)(D-1)/D$. As $N \to \infty$, $v_c \to 1$ similarly to the type I. For fixed $N$, on the other hand, $v_c \to 0$ as $D \to \infty$. This implies that, in the limit of $D \to \infty$, a finite number of local observables and a finite Schmidt rank of entangled state violate the type-II chained Bell inequality for $v_c\to 0.$

\section{Remarks}
\label{conclusion}
 In this work we have  presented a sufficient condition $\eqref{eq:ncfur}$ to certify violation  of a Bell-type inequality against white noise~\ref{whiten} for $v_c\to 0$. To this end, we define a set of quasi-separations (type II)  between observables. The set of quasi-separations do not obey permutation symmetry but satisfy triangle inequalities. Our formalism is quite general as this is based on violation of a Bell-type inequality, which is obtained by summing a set of elementary triangle inequalities. 
We show that  pure quantum states with Schmidt rank much lower than the dimension of white noise  are optimal to show the violations of the Bell-type inequality against white noise for infinitesimal values of $v_c.$ However, if the dimension of white noise matches with the Schmidt rank of the pure quantum states, then critical visibilities $v_c$ tend to  increase (see Fig.~\ref{fig:1}) with the  Schmidt rank of the pure quantum state. We extend our analysis for a chained Bell-type inequality with many measurement settings per site. 

Also, using another set of quasi-separations (type I) we obtain CGLMP inequality. Note that, as for the CGLMP inequality the sufficient condition~$\eqref{eq:ncfur}$ is not satisfied, we cannot obtain violations of the inequality against white noise for $v_c\to0.$ We also check the violation of the chained CGLMP-type inequality in which each of the two parties measures $N$ local observables.

In conclusion, one of the main results of this manuscript shows that, in a device-independent scenario, considering  critical visibility as a measure of resilience of  the violation of the Bell-type inequality against white noise may lead to inconclusive result if we do not have the knowledge of the Schmidt rank of the pure quantum states.

\section*{ACKNOWLEDGMENTS}
AD and JL would like to thank Marek Zukowski for stimulating discussions.
A.D. is supported by a KIAS Individual Grant (CG066401). J.K. is supported in part by KIAS Advanced Research Program
(No. CG014604). JL acknowledges the support of the National Research Foundation of Korea(NRF) grant
 funded by the Korea government(MSIT) (No. 2019R1A2C2005504 and No. 2019M3E4A1079666).

\appendix

\section{Quasi-separations and their triangle inequalities in local realistic models} 
\label{sec:sqdlrm}

A quasi-distance $d(x,y)$ between two points $x$ and $y$ satisfies (a) the non-negativity $d(x,y) \ge 0$ for every pair $(x,y)$, where the equality holds if and only if $x=y$, and (b) the triangle inequality (or subadditivity) 
\begin{eqnarray}
\label{eq:triang}
d(x,y) + d(y,z) \ge d(x,z)
\end{eqnarray}
for every triple $(x,y,z)$. On the other hand, a quasi-distance does not obey the permutation symmetry, $d(x,y) \ne d(y,x)$. This requires to mind the order of points in the triangle inequalities.

We prove that the statistical quasi-separation, as in Eq.~\eqref{eq:sqd}, is a quasi-distance in local realistic models. Every local realistic model assumes the existence of full joint probability distribution for all the local observables~\cite{Fine82}: In our work, $p(\{a_i,b_j\}) := p(a_1, a_3, \ldots, a_{2N-1}, b_2, b_4, \ldots, b_{2N})$ exists. Marginal probability distributions come from the full joint. The statistical quasi-separation $S(X,Y)$ between every two local observables $X$ and $Y$ in Eq.~\eqref{eq:sqd}, is now given by
\begin{equation}
\label{eq:sqdap}
\begin{aligned}
S(X, Y) &=\sum_{x,y} d(x, y)\, p(x, y| X, Y) \\
&= \sum_{\{a_i,b_j\}} d(x,y)\, p(\{a_i,b_j\}),
\end{aligned}
\end{equation}
where the second equality holds as the marginal $p(x, y| X, Y)$ results from the full joint $p(\{a_i,b_j\})$, $\{a_i,b_j\}$ stands for the outcome set of all the local observables, and $\{x,y\}$ is of the observable pair $X$ and $Y$ among them. Then, the quasi-separation $S(X,Y)$ is a quasi-distance: (a) $S(X,Y)$ is nonnegative for every pair of local observables $X$ and $Y$, as it is an average of nonnegative quasi-distance $d(x,y)$. (b) $S(X,Y) + S(Y,Z) - S(X,Z) \ge 0$ for every triple of local observables $(X,Y,Z)$, implying the triangle inequality. This is the case as it is a mere average of the nonnegative quantities, $d(x,y)+d(y,z)-d(x,z) \ge 0, \forall (x,y,z)$, over the full joint probability $p(\{a_i,b_j\})$ or the marginal $p(x,y,z)$. In other words, for every triple of local observables $(X,Y,Z)$ with outcomes $(x,y,z)$,
\begin{eqnarray}
\label{eq:sqdapta}
\sum_{\{a_i,b_j\}} p(\{a_i,b_j\}) \, \left( d(x, y) + d(y,z) - d(x,z) \right) \ge 0.
\end{eqnarray}
This equals the triangle inequality~\eqref{eq:sqdta}, 
\begin{equation}
S(X, Y) + S(Y,Z) \ge S(X,Z).  \tag{\ref{eq:sqdta}}
\end{equation}
We note that the existence of full joint probability bases the proof of triangle inequalities. To the contrary, such a full joint probability can not exist generally in quantum model.

These results remain valid for every quasi-distance $d(x,y)$, when `points' $x$ and $y$ are mapped to the outcomes of local observables $X$ and $Y$, respectively. Some typical forms of quasi-distances are employed in this work.

\section{Quasi-distances}
\label{sec:qdfbi}
We consider a candidate for quasi-distances $d(x,y)$ for Bell inequalities. It is given by
\begin{equation}
d(x,y) = \left[ f(x) - f(y) \right]_D, \tag{\ref{eq:qdcglmp1}}
\end{equation}
where $[x]_D = x \mod D$ is congruent to a nonnegative residue modulo $D$ for integral $x$, so that it belongs to the set $\{0, 1, ..., D-1\}$. Here, $f(x)$ is an integer-valued function of integral $x$. Every candidate in a form of Eq.~\eqref{eq:qdcglmp1} with an arbitrary function $f(x)$ is a quasi-distance, regardless whether $f(x)$ is linear or nonlinear. The candidate is nonnegative by definition. It satisfies the subadditivity, originating from the nonnegative residues $[x]_D$.

The nonnegative residues $[x]_D$ satisfy the subadditivity (or triangle inequality), i.e.
\begin{eqnarray}
\label{eq:nnrs}
[x+y]_D \le [x]_D + [y]_D.
\end{eqnarray}
This implies the reverse triangle inequality,
\begin{eqnarray}
[x]_D - [y]_D \le [x-y]_D.
\end{eqnarray}
The inverse of addition is given by
\begin{eqnarray}
[-x]_D = D -1 - [x-1]_D.
\end{eqnarray}
The subadditivity in Eq.~\eqref{eq:nnrs} proves that of the quasi-distance in Eq.~\eqref{eq:triang}. In other words, by Eq.~\eqref{eq:nnrs},
\begin{equation}
\left[ f(x) - f(y) \right]_D + \left[ f(y) - f(z) \right]_D \ge \left[ f(x) - f(z) \right]_D.
\end{equation}
This implies
\begin{equation}
d(x,y) + d(y,z) \ge d(x,z).
\end{equation}

\subsection{Quasi-separations of white noise}
\label{sec:qswn}

We prove that the quasi-separation $S_r$ for the state of white noise is given by
\begin{equation}
\label{eq:qsfwn}
\begin{aligned}
S_r	&= \frac{1}{2D} \left[ \left( \sum_{f=0}^{D-1} n(f) \right)^2 - \sum_{f=0}^{D-1}n^2(f) \right],
\end{aligned}
\end{equation}
where $n(f)$ is the number of $x$'s such that $f(x) = f$ for each integer $f$, i.e. 
\begin{equation}
n(f) = \sum_{x=0}^{D-1} \delta_{f(x), f}. 
\end{equation}
By definition, the quasi-separation for the distribution of white noise is given,
\begin{equation}
S_r = \frac{1}{D^2} \sum_{x,y=0}^{D-1} d(x,y) = \frac{1}{D^2} \sum_{x,y=0}^{D-1} \left[ f(x) - f(y) \right]_D.
\end{equation}
This can be written in terms of $n(f)$,
\begin{equation}
\begin{aligned}
S_r 	&= \frac{1}{D^2} \sum_{f_0,f_1=0}^{D-1} n(f_0) n(f_1) \left[f_0 - f_1 \right]_D.
\end{aligned}
\end{equation}
Note that
\begin{equation}
\left[ f_0 - f_1 \right]_D =
\left\{
\begin{aligned}
& f_0 - f_1 && \text{if $f_0 \ge f_1$} \\
& D - \left( f_1 - f_0 \right) && \text{if $f_0 < f_1$}
\end{aligned},
\right.
\end{equation}
and
\begin{equation}
\sum_{f_0,f_1} n(f_0) n(f_1) f_0 = \sum_{f_0,f_1} n(f_0) n(f_1) f_1.
\end{equation}
Then,
\begin{equation}
\begin{aligned}
S_r	&= \frac{1}{D} \sum_{f_0 < f_1}^{D-1} n(f_0) n(f_1) \\
	&= \frac{1}{2D} \left[ \left( \sum_{f=0}^{D-1} n(f) \right)^2 - \sum_{f=0}^{D-1}n^2(f) \right],
\end{aligned}
\end{equation}
where we used
\begin{equation*}
\begin{aligned}
\sum_{f_0 < f_1} n(f_0) n(f_1) = \sum_{f_0 > f_1} n(f_0) n(f_1).
\end{aligned}
\end{equation*}

Consider two typical forms of $f(x)$: (a) A linear function, $f_\text{I}(x) = x$ in Eq.~\eqref{eq:lfinnr}, and (b) a nonlinear function, 
$f_\text{II}(x) = \delta_{x,0}$ in Eq.~\eqref{eq:nlfinnr},
where $\delta_{x,y}$ is the Kronecker delta. The quasi-separations $S_r$ and $\text{SSR} = S_r/S_\text{max}$ are obtained. In case of (a) with $f(x) = f_\text{I}(x)$, $n(f) = 1$ for each $f$ and $S_\text{max} = D-1$. Eq.~\eqref{eq:qsfwn} is led to
\begin{align}
S_r = \frac{D-1}{2} \quad \text{and} \quad \text{SSR} = \frac{S_r}{S_\text{max}} = \frac{1}{2}.
\end{align}
In case of (b) with $f(x) = f_\text{II}(x)$, $n(0) = D-1$, $n(1) = 1$, $n(f > 1) = 0$, and $S_\text{max} = D-1$. By Eq.~\eqref{eq:qsfwn}, 
\begin{align}
S_r = \frac{D-1}{D} \quad \text{and} \quad \text{SSR} = \frac{S_r}{S_\text{max}} = \frac{1}{D}.
\end{align}
It is seen that $\text{SSR} \to 0$ as $D \to \infty$. 

\section{Quantum Bell functions}
\label{sec:qprob}

The input state $\hat{\rho} = |\psi\rangle \langle \psi|$ is assumed entangled with Schmidt rank $R \le D$, i.e. 
\begin{eqnarray}
|\psi\rangle =\frac{1}{\sqrt{R}}\sum_{j=0}^{R-1}|j j \rangle,
\end{eqnarray}
where $\{|j\rangle\}$ is a standard basis. If Schmidt rank $R$ is less than $D$, we decompose Hilbert space ${\cal H}$ into two subspaces ${\cal H}_R \oplus {\cal H}_R^\perp$, where ${\cal H}_R$ is spanned by $\{|j\rangle\}_{j=0}^{R-1}$ and its orthogonal complement ${\cal H}_R^\perp$ is by $\{|j\rangle\}_{i=R}^{D-1}$. A basis set of each local measurement is also divided into two subsets; one of Fourier transformed states and the other of the standard-basis states. More precisely, the measurement basis $\{|a\rangle_n\}$ of Alice's local observable $A_n$ is given by
\begin{equation}
\label{eq:lmbarr}
\ket{a}_n =
\left\{
\begin{aligned}
& \frac{1}{\sqrt{R}}\sum_{j=0}^{R-1} \omega_R^{j (a +\alpha_n)}|j \rangle && \text{if $a < R$} \\
& |a\rangle && \text{otherwise},
\end{aligned}
\right.
\end{equation}
where $\omega_R = \exp \left( 2i\pi / R \right)$ is a primitive $R$-th root of unity, and $\alpha_n$ is a constant to be chosen. 
Similarly, the measurement basis $\{|b\rangle_m\}$ of $B_m$ is given by 
\begin{eqnarray}
\label{eq:lmbbrr}
\ket{b}_m =
\left\{
\begin{aligned}
& \frac{1}{\sqrt{R}}\sum_{j=0}^{R-1} \omega_R^{-j (b +\beta_m)}|j \rangle && \text{if $b < R$} \\
& |b\rangle && \text{otherwise},
\end{aligned}
\right.
\end{eqnarray}
where $\beta_m$ is a constant to be chosen. We note that, if $R=D$, the bases are given by the upper terms in Eqs.~\eqref{eq:lmbarr} and \eqref{eq:lmbbrr}, respectively. The constants $\alpha_n$ and $\beta_m$ are chosen as
\begin{subequations}
\label{eq:const}
\begin{align}
\alpha_{2n-1}& = -\frac{2(n-1)}{2N} = -\frac{n-1}{N} \\
\beta_{2n} &= -\frac{2n-1}{2N}
\end{align}
\end{subequations}
for $1 \le n \le N$, where $N$ is the number of local observables at each site.
If each observer employs two local measurements, i.e. $N=2$, this is reduced to 
\begin{equation}
\begin{aligned}
& \alpha_1 = 0, & \beta_2 = -1/4, \\
& \alpha_3 = -1/2, & \beta_4 = -3/4.
\end{aligned}
\end{equation}

\subsection{Quantum probabilities}
\label{sec:qpampc}

The joint probability $p(a, b | A_n, B_m)$ that Alice and Bob respectively obtain outcomes $a$ and $b$ in their local measurements of $A_n$ and $B_m$, is given by $p_{nm}(a, b) := p(a, b|A_n, B_m) = \Tr \left( \hat{\rho} \, \hat{A}_n(a) \otimes \hat{B}_m(b) \right)$, where the projectors $\hat{A}_n(a) = |a\rangle_{n n} \langle a|$ and $\hat{B}_m(b) = |b\rangle_{m m}\langle b|$. Then,
\begin{eqnarray}
\label{probcglmprr}
p_{nm}(a, b) = 
\left\{ 
\begin{aligned}
& \frac{g_{1}^{n m}(0)}{g_{R}^{n m}(a-b)} && \text{if $a,b<R$} \\
&0 && \text{otherwise}
\end{aligned}
\right.
\end{eqnarray}
where  $g_{X}^{n m}(x)=X^3\sin^2\left[(\pi/X) (x+\alpha_n-\beta_m)\right]$. We note that $g_{X}^{nm}(x)$ is periodic with $X$, i.e. $g_{X}^{nm}(x) = g_{X}^{nm}(X+x)$. In particular, $g_1^{nm}$ is a constant function: $g_{1}^{nm}(x) = g_{1}^{nm}(0)$. If $R=D$, the probability is given by the upper term in Eq.~\eqref{probcglmprr}. From Eqs.~\eqref{eq:const}, we obtain
\begin{subequations}
\label{eq:gnmr}
\begin{align}
g^{2n-1, 2n}_X(x) &= g_X(x) \\
g^{2n+1, 2n}_X(x) &= g_X(-x) \\
g^{1,2N}_X(x) &= g_X(-x-1),
\end{align}
\end{subequations}
where $g_X(x) = X^3 \sin^2\left[\left(\pi/X\right)\left(x+1/2N\right)\right]$. For $X=1$, $g_1(x) = g_1(0) = \sin^2 \left( \pi / 2N\right)$. As the periodic function $g_R(x)$ is involved in the joint probability, it satisfies the normalization condition,
\begin{eqnarray}
\label{eq:bjp}
\begin{aligned}
& 1 = \sum_{a,b=0}^{R-1} \frac{g_{1}(0)}{g_{R}(a-b)} = R \sum_{x=0}^{R-1}  \frac{g_{1}(0)}{g_{R}(x)},
\end{aligned}
\end{eqnarray}
where we used the periodicity $g_X(x) = g_X(X+x)$ for the second equality. We introduce 
\begin{eqnarray}
\label{eq:mpcv}
\begin{aligned}
p_R(c):=R \, \frac{g_{1}(0)}{g_R(c)} = \frac{\sin^2\left( \pi/2N \right)}{R^2 \sin^2\left[ (\pi/R) (c+1/2N) \right]}
\end{aligned}
\end{eqnarray}
to be a marginal probability of the correlation variable $C$. Note $p_R(c)$ is periodic with $R$. The correlation variables $C = \left[A - B \right]_R = A - B \mod R$, due to the periodicity of $p_R(c)$ and the relation between Eqs.~\eqref{eq:bjp} and \eqref{eq:mpcv}. Note the modulus is $R$ whereas it is $D$ in the quasi-distance $d(x,y)$ of Eq.~\eqref{eq:qdcglmp1}. The average of $C$ is defined by
\begin{eqnarray}
\label{eq:ampcv}
\begin{aligned}
\bar{C}_R = \sum_{c=0}^{R-1} c \, p_R(c).
\end{aligned}
\end{eqnarray}

\subsection{Quantum Bell function of type I}
\label{sec:qbfti}

The quasi-separations $S_\text{I}(A_n,B_m)$ of type I is defined by $S_\text{I}(A_n,B_m) = \sum_{a,b} d(a,b) p_{nm}(a,b)$ with $d(a,b) = [a - b]_D$.
From Eqs.~\eqref{probcglmprr} and \eqref{eq:gnmr}, the quasi-separations $S_\text{I}(A_n,B_m)$ for $R=D$ are given as
\begin{subequations}
\begin{align}
S_\text{I}(A_{2n-1}, B_{2n}) 	&= S_\text{I}(B_{2n},A_{2n+1}) = \bar{C}_D, \\
S_\text{I}(A_1, B_{2N}) 		&= D-1 - \bar{C}_D,
\end{align}
\end{subequations}
where the first equation holds for each $n$ and $\bar{C}_D$ is given in Eq.~\eqref{eq:ampcv}.
Then the Bell function $I_{Nq}:=I_N(\hat{\rho})$ of type I is given by Eq.~\eqref{eq6}
\begin{eqnarray}
\begin{aligned}
I_{Nq} 	&= - (D-1) \, \xi, \\
\xi 		&= 1 - \frac{2N}{D-1} \bar{C}_D.
\end{aligned}
\end{eqnarray}
For white noise, $I_{Nr} = (N-1) (D-1)$. The critical visibility $v_c$ is given,
\begin{align}
v_c = \frac{1}{1 + \xi /(N-1)}.
\end{align}

\subsection{Quantum Bell function of type II}
\label{sec:qbftii}

The quasi-separations $S_\text{II}(A_n,B_m)$ of type II is defined by $S_\text{II}(A_n,B_m) = \sum_{a,b} d(a,b) p_{nm}(a,b)$ with $d(a,b) = [\delta_{a,0} - \delta_{b,0}]_D$.
From Eqs.~\eqref{probcglmprr} and \eqref{eq:gnmr}, the quasi-separations $S_\text{II}(A_n,B_m)$ are given as, for each $n$,
\begin{subequations}
\begin{align}
S_\text{II}(A_{2n-1}, B_{2n}) 	&= \frac{D}{R} \left(  1- p_R(0) \right), \\
S_\text{II}(B_{2n}, A_{2n+1}) 	&= \frac{D}{R} \left(  1- p_R(0) \right)	, \\
S_\text{II}(A_{1}, B_{2N}) 		&= \frac{D}{R} \left(  1- p_R(-1) \right),
\end{align}
\end{subequations}
where $p_R(c)$ is given in  Eq.~\eqref{eq:mpcv}.
The Bell function $I_{Nq} := I_N(\hat{\rho})$ is then given by Eq.~\eqref{eq6},
\begin{equation}
\label{eq:qbftiin}
\begin{aligned}
I_{Nq} &= - D \, \xi'_R, \\
\xi'_R &= \frac{1}{R} \left[ \left(1-p_R(-1)\right) - (2N-1) \left(1-p_{R}(0)\right) \right].
\end{aligned}
\end{equation}
As $I_{Nr} = 2(N-1)(D-1)/D$, the critical visibility $v_c$ is given by Eq.~\eqref{eq5c},
\begin{eqnarray}
\label{eq:qcvtiin}
v_c = \frac{1}{1 + \frac{D^2}{2(D-1)} \frac{1}{N-1}\, \xi'_R}.
\end{eqnarray}


\begin{thebibliography}{34}
\expandafter\ifx\csname natexlab\endcsname\relax\def\natexlab#1{#1}\fi
\expandafter\ifx\csname bibnamefont\endcsname\relax
  \def\bibnamefont#1{#1}\fi
\expandafter\ifx\csname bibfnamefont\endcsname\relax
  \def\bibfnamefont#1{#1}\fi
\expandafter\ifx\csname citenamefont\endcsname\relax
  \def\citenamefont#1{#1}\fi
\expandafter\ifx\csname url\endcsname\relax
  \def\url#1{\texttt{#1}}\fi
\expandafter\ifx\csname urlprefix\endcsname\relax\def\urlprefix{URL }\fi
\providecommand{\bibinfo}[2]{#2}
\providecommand{\eprint}[2][]{\url{#2}}

\bibitem[{\citenamefont{Bell}(1964)}]{Bell64}
\bibinfo{author}{\bibfnamefont{J.~S.} \bibnamefont{Bell}},
  \bibinfo{journal}{Physics} \textbf{\bibinfo{volume}{1}}, \bibinfo{pages}{195}
  (\bibinfo{year}{1964}).

\bibitem[{\citenamefont{Einstein et~al.}(1935)\citenamefont{Einstein, Podolsky,
  and Rosen}}]{EPR35}
\bibinfo{author}{\bibfnamefont{A.}~\bibnamefont{Einstein}},
  \bibinfo{author}{\bibfnamefont{B.}~\bibnamefont{Podolsky}}, \bibnamefont{and}
  \bibinfo{author}{\bibfnamefont{N.}~\bibnamefont{Rosen}},
  \bibinfo{journal}{Phys. Rev.} \textbf{\bibinfo{volume}{47}},
  \bibinfo{pages}{777} (\bibinfo{year}{1935}).

\bibitem[{\citenamefont{Polozova and Strauch}(2016)}]{polozova2016higher}
\bibinfo{author}{\bibfnamefont{E.}~\bibnamefont{Polozova}} \bibnamefont{and}
  \bibinfo{author}{\bibfnamefont{F.~W.} \bibnamefont{Strauch}},
  \bibinfo{journal}{Physical Review A} \textbf{\bibinfo{volume}{93}},
  \bibinfo{pages}{032130} (\bibinfo{year}{2016}).

\bibitem[{\citenamefont{Wu and Lidar}(2004)}]{Wu2004}
\bibinfo{author}{\bibfnamefont{L.-A.} \bibnamefont{Wu}} \bibnamefont{and}
  \bibinfo{author}{\bibfnamefont{D.~A.} \bibnamefont{Lidar}},
  \bibinfo{journal}{Physical Review A} \textbf{\bibinfo{volume}{70}},
  \bibinfo{pages}{062310} (\bibinfo{year}{2004}).

\bibitem[{\citenamefont{Kaszlikowski et~al.}(2000)\citenamefont{Kaszlikowski,
  Gnaci{\'n}ski, {\.Z}ukowski, Miklaszewski, and Zeilinger}}]{Kaszlikowski00}
\bibinfo{author}{\bibfnamefont{D.}~\bibnamefont{Kaszlikowski}},
  \bibinfo{author}{\bibfnamefont{P.}~\bibnamefont{Gnaci{\'n}ski}},
  \bibinfo{author}{\bibfnamefont{M.}~\bibnamefont{{\.Z}ukowski}},
  \bibinfo{author}{\bibfnamefont{W.}~\bibnamefont{Miklaszewski}},
  \bibnamefont{and}
  \bibinfo{author}{\bibfnamefont{A.}~\bibnamefont{Zeilinger}},
  \bibinfo{journal}{Physical Review Letters} \textbf{\bibinfo{volume}{85}},
  \bibinfo{pages}{4418} (\bibinfo{year}{2000}).

\bibitem[{\citenamefont{Collins et~al.}(2002)\citenamefont{Collins, Gisin,
  Linden, Massar, and Popescu}}]{CGLMP02}
\bibinfo{author}{\bibfnamefont{D.}~\bibnamefont{Collins}},
  \bibinfo{author}{\bibfnamefont{N.}~\bibnamefont{Gisin}},
  \bibinfo{author}{\bibfnamefont{N.}~\bibnamefont{Linden}},
  \bibinfo{author}{\bibfnamefont{S.}~\bibnamefont{Massar}}, \bibnamefont{and}
  \bibinfo{author}{\bibfnamefont{S.}~\bibnamefont{Popescu}},
  \bibinfo{journal}{Phys. Rev. Lett.} \textbf{\bibinfo{volume}{88}},
  \bibinfo{pages}{040404} (\bibinfo{year}{2002}).

\bibitem[{\citenamefont{Nielsen and Chuang}(2000)}]{Nielsen00}
\bibinfo{author}{\bibfnamefont{M.}~\bibnamefont{Nielsen}} \bibnamefont{and}
  \bibinfo{author}{\bibfnamefont{I.}~\bibnamefont{Chuang}},
  \emph{\bibinfo{title}{{Quantum Computation and Quantum Information}}}
  (\bibinfo{publisher}{Cambridge Univ. Press, Cambridge},
  \bibinfo{year}{2000}).

\bibitem[{\citenamefont{Loura et~al.}(2014)\citenamefont{Loura, Almeida, Andre,
  Pinto, Mateus, and Paunkovic}}]{loura2014}
\bibinfo{author}{\bibfnamefont{R.}~\bibnamefont{Loura}},
  \bibinfo{author}{\bibfnamefont{A.~J.} \bibnamefont{Almeida}},
  \bibinfo{author}{\bibfnamefont{P.~S.} \bibnamefont{Andre}},
  \bibinfo{author}{\bibfnamefont{A.~N.} \bibnamefont{Pinto}},
  \bibinfo{author}{\bibfnamefont{P.}~\bibnamefont{Mateus}}, \bibnamefont{and}
  \bibinfo{author}{\bibfnamefont{N.}~\bibnamefont{Paunkovic}},
  \bibinfo{journal}{Physical Review A} \textbf{\bibinfo{volume}{89}},
  \bibinfo{pages}{052336} (\bibinfo{year}{2014}).

\bibitem[{\citenamefont{Werner}(Oct 1989)}]{Werner89}
\bibinfo{author}{\bibfnamefont{R.~F.} \bibnamefont{Werner}},
  \bibinfo{journal}{Physical Review A} \textbf{\bibinfo{volume}{40}},
  \bibinfo{pages}{4277} (\bibinfo{year}{Oct 1989}).

\bibitem[{\citenamefont{Clauser et~al.}(1969)\citenamefont{Clauser, Horne,
  Shimony, and Holt}}]{CHSH69}
\bibinfo{author}{\bibfnamefont{J.~F.} \bibnamefont{Clauser}},
  \bibinfo{author}{\bibfnamefont{M.~A.} \bibnamefont{Horne}},
  \bibinfo{author}{\bibfnamefont{A.}~\bibnamefont{Shimony}}, \bibnamefont{and}
  \bibinfo{author}{\bibfnamefont{R.~A.} \bibnamefont{Holt}},
  \bibinfo{journal}{Phys. Rev. Lett.} \textbf{\bibinfo{volume}{23}},
  \bibinfo{pages}{880} (\bibinfo{year}{1969}).

\bibitem[{\citenamefont{Clauser and Horne}(1974)}]{Clauser74}
\bibinfo{author}{\bibfnamefont{J.~F.} \bibnamefont{Clauser}} \bibnamefont{and}
  \bibinfo{author}{\bibfnamefont{M.~A.} \bibnamefont{Horne}},
  \bibinfo{journal}{Phys. Rev. D} \textbf{\bibinfo{volume}{10}},
  \bibinfo{pages}{526} (\bibinfo{year}{1974}).

\bibitem[{\citenamefont{Lim et~al.}(2010)\citenamefont{Lim, Ryu, Yoo, Lee,
  Bang, and Lee}}]{lim2010genuinely}
\bibinfo{author}{\bibfnamefont{J.}~\bibnamefont{Lim}},
  \bibinfo{author}{\bibfnamefont{J.}~\bibnamefont{Ryu}},
  \bibinfo{author}{\bibfnamefont{S.}~\bibnamefont{Yoo}},
  \bibinfo{author}{\bibfnamefont{C.}~\bibnamefont{Lee}},
  \bibinfo{author}{\bibfnamefont{J.}~\bibnamefont{Bang}}, \bibnamefont{and}
  \bibinfo{author}{\bibfnamefont{J.}~\bibnamefont{Lee}}, \bibinfo{journal}{New
  Journal of Physics} \textbf{\bibinfo{volume}{12}}, \bibinfo{pages}{103012}
  (\bibinfo{year}{2010}).

\bibitem[{\citenamefont{Peres}(1999)}]{Peres99}
\bibinfo{author}{\bibfnamefont{A.}~\bibnamefont{Peres}},
  \bibinfo{journal}{Found. Phys.} \textbf{\bibinfo{volume}{29}},
  \bibinfo{pages}{589} (\bibinfo{year}{1999}).

\bibitem[{\citenamefont{Santos}(1986)}]{santos1986}
\bibinfo{author}{\bibfnamefont{E.}~\bibnamefont{Santos}},
  \bibinfo{journal}{Physics letters A} \textbf{\bibinfo{volume}{115}},
  \bibinfo{pages}{363} (\bibinfo{year}{1986}).

\bibitem[{\citenamefont{Zohren and Gill}(2008)}]{zohren2008}
\bibinfo{author}{\bibfnamefont{S.}~\bibnamefont{Zohren}} \bibnamefont{and}
  \bibinfo{author}{\bibfnamefont{R.~D.} \bibnamefont{Gill}},
  \bibinfo{journal}{Physical review letters} \textbf{\bibinfo{volume}{100}},
  \bibinfo{pages}{120406} (\bibinfo{year}{2008}).

\bibitem[{\citenamefont{Pearle}(1970)}]{Pearle1970}
\bibinfo{author}{\bibfnamefont{P.~M.} \bibnamefont{Pearle}},
  \bibinfo{journal}{Physical Review D} \textbf{\bibinfo{volume}{2}},
  \bibinfo{pages}{1418} (\bibinfo{year}{1970}).

\bibitem[{\citenamefont{Pykacz and Santos}(1991)}]{pykacz1991}
\bibinfo{author}{\bibfnamefont{J.}~\bibnamefont{Pykacz}} \bibnamefont{and}
  \bibinfo{author}{\bibfnamefont{E.}~\bibnamefont{Santos}},
  \bibinfo{journal}{Journal of mathematical physics}
  \textbf{\bibinfo{volume}{32}}, \bibinfo{pages}{1287} (\bibinfo{year}{1991}).

\bibitem[{\citenamefont{{\.Z}ukowski and Dutta}(2014)}]{Zukowski2014}
\bibinfo{author}{\bibfnamefont{M.}~\bibnamefont{{\.Z}ukowski}}
  \bibnamefont{and} \bibinfo{author}{\bibfnamefont{A.}~\bibnamefont{Dutta}},
  \bibinfo{journal}{Physical Review A} \textbf{\bibinfo{volume}{90}},
  \bibinfo{pages}{012106} (\bibinfo{year}{2014}).

\bibitem[{\citenamefont{Greenberger et~al.}(1989)\citenamefont{Greenberger,
  Horne, and Zeilinger}}]{GHZ89}
\bibinfo{author}{\bibfnamefont{D.~M.} \bibnamefont{Greenberger}},
  \bibinfo{author}{\bibfnamefont{M.~A.} \bibnamefont{Horne}}, \bibnamefont{and}
  \bibinfo{author}{\bibfnamefont{A.}~\bibnamefont{Zeilinger}}, in
  \emph{\bibinfo{booktitle}{Bell's Theorem, Quantum Theory, and Conceptions of
  the Universe}}, edited by
  \bibinfo{editor}{\bibfnamefont{M.}~\bibnamefont{Kafatos}}
  (\bibinfo{publisher}{Kluwer, Dordrecht}, \bibinfo{year}{1989}).

\bibitem[{\citenamefont{Dutta et~al.}(2018)\citenamefont{Dutta, Nahm, Lee, and
  {\.Z}ukowski}}]{Dutta2018}
\bibinfo{author}{\bibfnamefont{A.}~\bibnamefont{Dutta}},
  \bibinfo{author}{\bibfnamefont{T.-U.} \bibnamefont{Nahm}},
  \bibinfo{author}{\bibfnamefont{J.}~\bibnamefont{Lee}}, \bibnamefont{and}
  \bibinfo{author}{\bibfnamefont{M.}~\bibnamefont{{\.Z}ukowski}},
  \bibinfo{journal}{New Journal of Physics} \textbf{\bibinfo{volume}{20}},
  \bibinfo{pages}{093006} (\bibinfo{year}{2018}).

\bibitem[{\citenamefont{Acin et~al.}(2002)\citenamefont{Acin, Durt, Gisin, and
  Latorre}}]{Acin2002}
\bibinfo{author}{\bibfnamefont{A.}~\bibnamefont{Acin}},
  \bibinfo{author}{\bibfnamefont{T.}~\bibnamefont{Durt}},
  \bibinfo{author}{\bibfnamefont{N.}~\bibnamefont{Gisin}}, \bibnamefont{and}
  \bibinfo{author}{\bibfnamefont{J.~I.} \bibnamefont{Latorre}},
  \bibinfo{journal}{Physical Review A} \textbf{\bibinfo{volume}{65}},
  \bibinfo{pages}{052325} (\bibinfo{year}{2002}).

\bibitem[{\citenamefont{Junge et~al.}(2010)\citenamefont{Junge, Palazuelos,
  P\'erez-Garc\'ia, Villanueva, and Wolf}}]{Junge2010}
\bibinfo{author}{\bibfnamefont{M.}~\bibnamefont{Junge}},
  \bibinfo{author}{\bibfnamefont{C.}~\bibnamefont{Palazuelos}},
  \bibinfo{author}{\bibfnamefont{D.}~\bibnamefont{P\'erez-Garc\'ia}},
  \bibinfo{author}{\bibfnamefont{I.}~\bibnamefont{Villanueva}},
  \bibnamefont{and} \bibinfo{author}{\bibfnamefont{M.~M.} \bibnamefont{Wolf}},
  \bibinfo{journal}{Communications in Mathematical Physics}
  \textbf{\bibinfo{volume}{300}}, \bibinfo{pages}{715–739}
  (\bibinfo{year}{2010}).

\bibitem[{\citenamefont{Dutta et~al.}(2016)\citenamefont{Dutta, Ryu, Laskowski,
  and {\.Z}ukowski}}]{dutta2016}
\bibinfo{author}{\bibfnamefont{A.}~\bibnamefont{Dutta}},
  \bibinfo{author}{\bibfnamefont{J.}~\bibnamefont{Ryu}},
  \bibinfo{author}{\bibfnamefont{W.}~\bibnamefont{Laskowski}},
  \bibnamefont{and}
  \bibinfo{author}{\bibfnamefont{M.}~\bibnamefont{{\.Z}ukowski}},
  \bibinfo{journal}{Physics Letters A} \textbf{\bibinfo{volume}{380}},
  \bibinfo{pages}{2191} (\bibinfo{year}{2016}).

\bibitem[{\citenamefont{Gruca et~al.}(2012)\citenamefont{Gruca, Laskowski, and
  {\.Z}ukowski}}]{Gruca2012}
\bibinfo{author}{\bibfnamefont{J.}~\bibnamefont{Gruca}},
  \bibinfo{author}{\bibfnamefont{W.}~\bibnamefont{Laskowski}},
  \bibnamefont{and}
  \bibinfo{author}{\bibfnamefont{M.}~\bibnamefont{{\.Z}ukowski}},
  \bibinfo{journal}{Physical Review A} \textbf{\bibinfo{volume}{85}},
  \bibinfo{pages}{022118} (\bibinfo{year}{2012}).

\bibitem[{\citenamefont{Khrennikov}(2008)}]{khrennikov2008bell}
\bibinfo{author}{\bibfnamefont{A.}~\bibnamefont{Khrennikov}},
  \bibinfo{journal}{Entropy} \textbf{\bibinfo{volume}{10}}, \bibinfo{pages}{19}
  (\bibinfo{year}{2008}).

\bibitem[{\citenamefont{Laskowski et~al.}(2015)\citenamefont{Laskowski,
  V{\'e}rtesi, and Wie{\'s}niak}}]{laskowski2015highly}
\bibinfo{author}{\bibfnamefont{W.}~\bibnamefont{Laskowski}},
  \bibinfo{author}{\bibfnamefont{T.}~\bibnamefont{V{\'e}rtesi}},
  \bibnamefont{and}
  \bibinfo{author}{\bibfnamefont{M.}~\bibnamefont{Wie{\'s}niak}},
  \bibinfo{journal}{Journal of Physics A: Mathematical and Theoretical}
  \textbf{\bibinfo{volume}{48}}, \bibinfo{pages}{465301}
  (\bibinfo{year}{2015}).

\bibitem[{\citenamefont{Salavrakos et~al.}(2016)\citenamefont{Salavrakos,
  Augusiak, Tura, Wittek, Ac{\'\i}n, and Pironio}}]{salavrakos2016bell}
\bibinfo{author}{\bibfnamefont{A.}~\bibnamefont{Salavrakos}},
  \bibinfo{author}{\bibfnamefont{R.}~\bibnamefont{Augusiak}},
  \bibinfo{author}{\bibfnamefont{J.}~\bibnamefont{Tura}},
  \bibinfo{author}{\bibfnamefont{P.}~\bibnamefont{Wittek}},
  \bibinfo{author}{\bibfnamefont{A.}~\bibnamefont{Ac{\'\i}n}},
  \bibnamefont{and} \bibinfo{author}{\bibfnamefont{S.}~\bibnamefont{Pironio}},
  \bibinfo{journal}{arXiv preprint arXiv:1607.04578}  (\bibinfo{year}{2016}).

\bibitem[{\citenamefont{Sen et~al.}(2003)\citenamefont{Sen, Sen, Wie{\'s}niak,
  Kaszlikowski, {\.Z}ukowski et~al.}}]{sen2003multiqubit}
\bibinfo{author}{\bibfnamefont{A.}~\bibnamefont{Sen}},
  \bibinfo{author}{\bibfnamefont{U.}~\bibnamefont{Sen}},
  \bibinfo{author}{\bibfnamefont{M.}~\bibnamefont{Wie{\'s}niak}},
  \bibinfo{author}{\bibfnamefont{D.}~\bibnamefont{Kaszlikowski}},
  \bibinfo{author}{\bibfnamefont{M.}~\bibnamefont{{\.Z}ukowski}},
  \bibnamefont{et~al.}, \bibinfo{journal}{Physical Review A}
  \textbf{\bibinfo{volume}{68}}, \bibinfo{pages}{062306}
  (\bibinfo{year}{2003}).

\bibitem[{\citenamefont{V{\'e}rtesi}(2008)}]{vertesi2008more}
\bibinfo{author}{\bibfnamefont{T.}~\bibnamefont{V{\'e}rtesi}},
  \bibinfo{journal}{Physical Review A} \textbf{\bibinfo{volume}{78}},
  \bibinfo{pages}{032112} (\bibinfo{year}{2008}).

\bibitem[{\citenamefont{Ac{\'\i}n et~al.}(2005)\citenamefont{Ac{\'\i}n, Gill,
  and Gisin}}]{acin2005}
\bibinfo{author}{\bibfnamefont{A.}~\bibnamefont{Ac{\'\i}n}},
  \bibinfo{author}{\bibfnamefont{R.}~\bibnamefont{Gill}}, \bibnamefont{and}
  \bibinfo{author}{\bibfnamefont{N.}~\bibnamefont{Gisin}},
  \bibinfo{journal}{Physical review letters} \textbf{\bibinfo{volume}{95}},
  \bibinfo{pages}{210402} (\bibinfo{year}{2005}).

\bibitem[{\citenamefont{Zeilinger et~al.}(1993)\citenamefont{Zeilinger,
  {\.Z}ukowski, Horne, Bernstein, and Greenberger}}]{zeilinger1993einstein}
\bibinfo{author}{\bibfnamefont{A.}~\bibnamefont{Zeilinger}},
  \bibinfo{author}{\bibfnamefont{M.}~\bibnamefont{{\.Z}ukowski}},
  \bibinfo{author}{\bibfnamefont{M.}~\bibnamefont{Horne}},
  \bibinfo{author}{\bibfnamefont{H.}~\bibnamefont{Bernstein}},
  \bibnamefont{and}
  \bibinfo{author}{\bibfnamefont{D.}~\bibnamefont{Greenberger}}, in
  \emph{\bibinfo{booktitle}{Fundamental aspects of quantum theory}}, edited by
  \bibinfo{editor}{\bibfnamefont{J.}~\bibnamefont{Anandan}} \bibnamefont{and}
  \bibinfo{editor}{\bibfnamefont{J.}~\bibnamefont{Safko}}
  (\bibinfo{publisher}{World Scientific}, \bibinfo{address}{Singapore},
  \bibinfo{year}{1993}).

\bibitem[{\citenamefont{Son et~al.}(2006)\citenamefont{Son, Lee, and
  Kim}}]{Son06a}
\bibinfo{author}{\bibfnamefont{W.}~\bibnamefont{Son}},
  \bibinfo{author}{\bibfnamefont{J.}~\bibnamefont{Lee}}, \bibnamefont{and}
  \bibinfo{author}{\bibfnamefont{M.~S.} \bibnamefont{Kim}},
  \bibinfo{journal}{Phys. Rev. Lett.} \textbf{\bibinfo{volume}{96}},
  \bibinfo{pages}{060406} (\bibinfo{year}{2006}).

\bibitem[{\citenamefont{Bancal et~al.}(2011)\citenamefont{Bancal, Brunner,
  Gisin, and Liang}}]{bancal2011detecting}
\bibinfo{author}{\bibfnamefont{J.-D.} \bibnamefont{Bancal}},
  \bibinfo{author}{\bibfnamefont{N.}~\bibnamefont{Brunner}},
  \bibinfo{author}{\bibfnamefont{N.}~\bibnamefont{Gisin}}, \bibnamefont{and}
  \bibinfo{author}{\bibfnamefont{Y.-C.} \bibnamefont{Liang}},
  \bibinfo{journal}{Physical review letters} \textbf{\bibinfo{volume}{106}},
  \bibinfo{pages}{020405} (\bibinfo{year}{2011}).

\bibitem[{\citenamefont{Fine}(1982)}]{Fine82}
\bibinfo{author}{\bibfnamefont{A.}~\bibnamefont{Fine}}, \bibinfo{journal}{Phys.
  Rev. Lett.} \textbf{\bibinfo{volume}{48}}, \bibinfo{pages}{291}
  (\bibinfo{year}{1982}).

\end{thebibliography}
\end{document}